\documentclass[preprint,12pt,5p,twocolumn]{elsarticle}

\usepackage{lineno}
\usepackage{amsmath}
\usepackage{amsthm}
\usepackage{hyperref}
\modulolinenumbers[5]

\usepackage{amssymb}

\journal{Physics Letters B}

\biboptions{sort&compress}

\begin{document}

\begin{frontmatter}
\title{Jet quenching as a probe of the initial stages in heavy-ion collisions\tnoteref{t1}}

\author[jlab]{Carlota Andres}
\ead{carlota@jlab.org}

\author[igfae]{N\'estor Armesto}
\ead{nestor.armesto@usc.es}

\author[jyv,hel]{Harri Niemi}
\ead{harri.m.niemi@jyu.fi}

\author[cern,hel]{Risto Paatelainen}
\ead{risto.sakari.paatelainen@cern.ch}

\author[igfae]{Carlos A. Salgado}
\ead{carlos.salgado@usc.es}

\address[jlab]{Jefferson Lab, 12000 Jefferson Avenue, Newport News, Virginia 23606, USA}
\address[igfae]{Instituto Galego de F\'isica de Altas Enerx\'ias IGFAE, Universidade de Santiago de Compostela, E-15782 Galicia-Spain}
\address[jyv]{University of Jyv\"askyl\"a, Department of Physics, P.O. Box 35, FI-40014 University of Jyv\"askyl\"a, Finland}
\address[hel]{Helsinki Institute of Physics, P.O. Box 64, FI-00014 University of Helsinki, Finland}
\address[cern]{Theoretical Physics Department, CERN, CH-1211 Gen\`eve 23, Switzerland}

\begin{abstract}
Jet quenching provides a very flexible variety of observables which are sensitive to different energy- and time-scales of the strongly interacting matter created in heavy-ion collisions. Exploiting this versatility would make jet quenching  an excellent chronometer of the yoctosecond structure of the evolution process. Here we show, for the first time, that a combination of jet quenching observables is sensitive to the initial stages of heavy-ion collisions, when the approach to local thermal equilibrium is expected to happen. Specifically, we find that in order to reproduce at the same time the inclusive particle production suppression, $R_{AA}$, and the high-$p_T$ azimuthal asymmetries,  $v_2$,  energy loss must be strongly suppressed for the first $\sim 0.6$ fm. This exploratory analysis shows the potential of jet observables, possibly more sophisticated than the ones studied here, to constrain the dynamics of the initial stages of the evolution.
\end{abstract}

\tnotetext[t1]{JLAB-THY-19-2888, CERN-TH-2019-012.}


\end{frontmatter}

\section{Introduction}

Heavy-ion collisions are the experimental tools designed to study the properties of 
the hot and dense Quark Gluon Plasma (QGP). After two decades of experiments at the Relativistic Heavy Ion Collider (RHIC) and the Large Hadron Collider (LHC), jet quenching, the modification of the Quantum Chromodynamics (QCD) jet structures due to their interaction with the surrounding matter, has become a fundamental tool for this program. Although the QGP is routinely produced and studied in these colliders, the actual process that so efficiently leads to the production of this locally thermalized state starting from a completely out-of-equilibrium collision system is largely unknown. This process must happen in a very short time, ${\cal O}(1 \,{\rm fm})$ or a few yoctoseconds. This is why this line of research, that has become one of the most active and interesting topics in QCD, is sometimes nicknamed {\it Initial Stages}. Up to now, all experimental information on the initial stages of the evolution comes, essentially, from azimuthal asymmetries in correlations between different particles in the soft regime (say, $p_T\lesssim 5$ GeV), and from deep inelastic scattering \cite{Mantysaari:2017cni,Mantysaari:2018zdd}.

Furthermore, recent experimental results from the LHC, and later from RHIC, in \textit{small system} p-Pb, high-multiplicity p-p and d-Au collisions, show characteristics \cite{Loizides:2016tew} usually attributed to QGP formation. Indeed, usual key probes of the QGP, such as long-range angular correlations and flow harmonics \cite{Khachatryan:2010gv,Abelev:2012ola,Adare:2013piz,Chatrchyan:2013nka,Abelev:2014mda,Aaboud:2017acw,Khachatryan:2015waa,Sirunyan:2018toe}, and the strangeness enhancement \cite{ALICE:2017jyt} have been observed in small systems. Interestingly, the only long-established QGP signature missing in these experimental data is jet quenching \cite{Khachatryan:2016odn}. Since thermalization and jet quenching are manifestations of basically the same dynamics, the presence of the former and the absence of the latter in these systems is surprising. For this reason, there is an ample consensus that jet quenching is critical to understand small systems and thermalization. We will argue here that jet quenching can be used, in fact, as a complementary and versatile way to probe the dynamics at the early times of the evolution. Actually, jets are extended objects in space and time, and different modifications measure different time or energy scales \cite{Andrews:2018jcm,Apolinario:2017sob}.

Using azimuthal asymmetries of hard particles as a jet quenching probe was proposed for the first time in \cite{Wang:2000fq,Gyulassy:2000gk}. The first data on high-$p_T$ elliptic flow, $v_2$, was published in 2006 by the PHENIX Collaboration \cite{Adler:2005rg}. However, even though the nuclear modification factor, $R_{AA}$, was fairly-well described by all the energy loss formalisms (e.g. embedded in event-by-event (EbyE) hydrodynamics \cite{Renk:2011qi}), the computed high-$p_T$ elliptic flow underestimated the experimental data \cite{Xu:2014ica}, an issue addressed in many studies \cite{Liao:2008dk,Jia:2011pi,Betz:2011tu,Armesto:2011ht,Betz:2012qq,Betz:2014cza,Xu:2014tda,Horowitz:2015dta,Xu:2015bbz,Adam:2015mda,Ramamurti:2017zjn} along the last decade. It was argued in \cite{Luzum:2012da,Noronha-Hostler:2016eow} that soft-hard correlations are decisive to properly determine the harmonic coefficients in the hard sector, whose correct definition is given by the scalar product, $v_n^{\rm SP}$  \cite{Noronha-Hostler:2016eow}, to be defined below.

In this work, we compute the azimuthally averaged $R_{AA}$ for the 20--30\% centrality class in $\sqrt{s_{\mathrm{NN}}}$ = 2.76 TeV Pb-Pb collisions at the LHC \cite{Abelev:2012hxa}. We have also checked  that our conclusions hold for other centrality classes, see Figs.~\ref{fig:T0-10}-\ref{fig:T40-50} in~\ref{appendix-sec1}. Our framework consists of a radiative energy loss implemented with the Quenching Weights (QWs) from Ref.~\cite{Salgado:2003gb}, embedded in an EKRT EbyE hydrodynamic simulation of the medium \cite{Niemi:2015qia}. Following the approach in \cite{Armesto:2009zi,Andres:2016iys}, we define the jet transport coefficient as $\hat{q} \equiv \,K\cdot 2\,\varepsilon^{3/4} $, driven by the ideal estimate $\hat{q}_{ideal }\sim 2\, \varepsilon^{3/4}$ \cite{Baier:2002tc}. The local energy density $\varepsilon$, is taken from EKRT hydrodynamic profiles, so that there is only one free parameter, the $K$-factor, which is fitted to the high-$p_T$ $R_{AA}$ experimental data \cite{Abelev:2012hxa} and used for the calculation of the high-$p_T$ harmonic coefficients.

We will show that the treatment of initial stages is crucial for the simultaneous description of both type of observables, since the jet harmonic coefficients show up to be very sensitive to the starting point of the quenching. In fact, the experimental data on $v_2$ at high-$p_T$ can only be described by delaying the beginning of the energy loss for $\sim 0.6$ fm. This general conclusion that we draw here for the first time\footnote{In \cite{Renk:2010qx} the authors comment that energy loss models with delayed quenching describe better in- and out-of-plane $R_{AA}$ data at RHIC, but no claim is made on the potential for constraining properties of the early stages.} is not limited to our specific implementation, since all studies that  describe the jet harmonic coefficients start the energy loss and hydrodynamical evolution at the same time \cite{Noronha-Hostler:2016eow, Betz:2016ayq,Zigic:2018ovr,Shi:2018lsf,Shi:2018vys}, implicitly implementing this time delay. We do not attempt here a comprehensive study of experimental data on $R_{AA}$ and $v_n$ but rather to show the importance of the initial stages of the evolution for their correct interpretation. It would be tempting, on the other hand, to relate our findings
to the absence of jet quenching in p-Pb collisions. We leave these studies for future works.

\section{The formalism}
\paragraph{Energy loss:} We follow the same formalism as in \cite{Andres:2016iys}, to which we refer the reader for further details. For a discussion on its limitations see also \cite{Armesto:2011ht}. Here we summarize its most relevant features. The cross section of a hadron $h$ at rapidity $y$ and transverse momentum $p_T$ is given by
\begin{multline}
\frac{d\sigma^{AA\to h}}{dydp_T} = \int dq_T\,dz\frac{d\sigma^{AA\to k}}{dydq_T}\,P(\epsilon)\, \\
\times  \,D_{k\to h}(z,\mu_F\equiv p_T)\,\delta\left(p_T-z(1-\epsilon)q_T\right),
\label{eq:crossec}
\end{multline}
where the cross section for producing a parton $k$,\ $d\sigma^{AA\to k}/dydq_T$, is computed at next-to leading order (NLO) by using the code in \cite{Stratmann:2001pb}. For the parton distribution functions, we use CTEQ6.6M \cite{Nadolsky:2008zw} together with EPS09 nuclear modifications \cite{Eskola:2009uj}. For the fragmentation functions (FFs) $D_{k\to h}(z,\mu_F)$, we use either DSS07 \cite{deFlorian:2007aj} or DSS14 \cite{deFlorian:2014xna}. The QWs $P(\epsilon)$ are employed in the multiple soft approximation \cite{Salgado:2003gb}\footnote{Our results and conclusions remain for scattering on a single center instead of multiple soft scatterings (see Fig.~\ref{fig:GLV} in~\ref{appendix-sec1}).}. These probability distributions depend on two variables, $\omega_c$ and $R$, which, for a dynamic expanding medium, are proportional, respectively, to the first and second moment of the jet quenching parameter $\hat{q}(\xi)$, defined along the trajectory of the radiating parton parametrized by $\xi$ \cite{Salgado:2003gb,Andres:2016iys}. Therefore, we only need a definition of the jet transport coefficient in terms of the local properties of the medium. We make use of the aforementioned expression\footnote{Other energy loss models that include  flow effects \cite{Noronha-Hostler:2016eow} require the same delayed quenching to describe the high-$p_T$ $v_n$.}:
\begin{equation}
 \hat{q}(\xi) \:= \:K \cdot 2\,\varepsilon^{3/4}(\xi).
 \label{eq:qhat}
\end{equation}

The previous equation is valid both for the partonic and for the hadronic phase of the evolution \cite{Baier:2002tc}. Nevertheless, most of the phenomenological works that try to extract the value of the quenching parameter assume no energy loss during the hadronic phase \cite{Burke:2013yra}. We analyze here two different scenarios: ending the energy loss at the chemical freeze-out $T_{\mathrm{q}} = T_{\mathrm{chem}}$ = 175 MeV, that is, no energy loss in the hadronic phase, and using Eq.~(\ref{eq:qhat}) all the way down to the kinetic freeze-out $T_{\mathrm{q}} = T_{\mathrm{dec}}$ = 100 MeV, i.e.,  including jet quenching in both phases\footnote{$T_{\mathrm{q}}$ denotes the temperature at which we stop the energy loss.}.

\paragraph{EKRT hydrodynamics:}
The EbyE fluctuating initial energy density profiles for the hydrodynamical evolution are calculated within the EKRT framework~\cite{Eskola:1999fc}. This framework is based on the collinearly factorized NLO computation in perturbative QCD (pQCD) of minijet transverse energy production and the conjecture of gluon saturation. The saturation momentum $p_{\rm sat}$ controls the computed transverse energy production, and is a function of the given collision energy $\sqrt{s_{\rm NN}}$, the nuclear mass number $A$, and its dependence on the transverse coordinate $\mathbf{x}_\perp$ comes through the product of the nuclear thickness functions $T_A(\mathbf{x}_\perp)$, computed event-by-event. The essential free parameter $K_{\rm sat}$ in the saturation conjecture is fixed by the charged hadron multiplicity in 0--5\% Pb-Pb collisions at $\sqrt{s_{\rm NN}} = 2.76$ TeV. Once $K_{\rm sat}$ is fixed, the initial energy density profiles can be computed for any $\sqrt{s_{\rm NN}}$ and $A$ as long as the saturation momentum remains in the perturbative regime, $p_{\rm sat} = p_{\rm sat}(\sqrt{s_{\rm NN}}, A, T_A T_A(\mathbf{x}_\perp)) > p_{\rm min} = 1$ GeV. The formation time of the initial condition is then obtained as $\tau_{\rm f} = 1/p_{\rm min} = 0.197$ fm.

After  formation, the subsequent spacetime evolution is computed using a boost-invariant transient Israel-Stewart type of second order relativistic dissipative hydrodynamics, where the essential physical inputs are the QCD matter equation of state and the temperature dependence of shear viscosity $\eta/s(T)$, for details see Ref.~\cite{Niemi:2015qia}. In particular, we obtain the spacetime evolution of the energy density profile $\varepsilon(\tau, \mathbf{x}_\perp)$ for each event, which are then used in the computation of the jet quenching parameter in Eq.~(\ref{eq:qhat}).

As an equation of state (EoS) we use the s95p parametrization of the lattice QCD results~\cite{Huovinen:2009yb} with chemical freeze-out implemented as in Ref.~\cite{Huovinen:2007xh}, and the shear viscosity parametrization is $\eta/s(T) = param1$ from Ref.~\cite{Niemi:2015qia}. The corresponding results for soft hadronic observables like multiplicity, average transverse momentum, flow coefficient and flow correlations are in an excellent agreement with the measurements of 200 GeV Au-Au collisions at RHIC, and 2.76 TeV Pb-Pb, 5.023 TeV Pb-Pb and 5.44 TeV Xe-Xe collisions at the LHC \cite{Niemi:2015qia, Niemi:2015voa, Eskola:2017bup, Niemi:2018ijm}.

\paragraph{Early-times treatment:} The dynamics prior to the applicability of hydrodynamics and, therefore, the associate energy loss phenomena, are not established yet. Thus, there is freedom in the definition of $\hat{q}(\xi)$ from the production time of the hadron to the initialization proper time $\tau_{\rm f}$ of EKRT EbyE hydrodynamics, see Eq.~(\ref{eq:qhat}). Energy loss in the BDMPS-Z formalism does not require, in principle, neither thermalization nor isotropization, so for times smaller than $\tau_{\rm f}$ it can be employed and $\hat{q}(\xi)$ has to be obtained via extrapolations. Up to now, any phenomenological study of this kind -- except explicitly indicated -- assumes no quenching during the early stages of the collision\footnote{See Refs.~\cite{Armesto:2009zi} and~\cite{Andres:2016iys} for some early time extrapolations.}. Indeed, all the proposed solutions to the long-standing problem of describing the high-$p_T$ $v_2$ delay the interaction of the hard parton with the medium up to the initial time of the hydrodynamic simulation \cite{Noronha-Hostler:2016eow, Betz:2016ayq,Zigic:2018ovr}, usually use $\tau_{\rm f}$ = 0.6 fm, or require a very substantial growth of $\hat{q}$ for temperatures close to the deconfinement temperature \cite{Shi:2018lsf,Shi:2018vys}. Since the starting time of EKRT EbyE hydrodynamics is set to $\tau_{\rm f}$ = 0.197 fm, we can study how the $R_{AA}$ and high-$p_T$ jet harmonic coefficients vary when we delay the jet quenching up to a time comparable with that in \cite{Noronha-Hostler:2016eow, Betz:2016ayq,Zigic:2018ovr}. Denoting by $\tau_{\rm q}$ the time where the jet quenching begins, we consider the following three cases:
\begin{itemize}
\item[i)] $\tau_{\rm q}=0$.
Here, $\hat{q}(\xi)=  \hat{q}(\tau_{\rm f})$ for $\xi < \tau_{\rm f} = 0.197$ fm.
\item[ii)] $\tau_{\rm q}=0.197$ fm. Here, $\hat{q}(\xi)=$ 0 for $\xi < \tau_{\rm f} = 0.197$ fm. In this case, the quenching begins at $0.197$ fm.
\item[iii)] $\tau_{\rm q}=0.572$ fm. Here, $\hat{q}(\xi)=$ 0 for $\xi < \tau_{\rm q} = 0.572$ fm. Hence, the energy loss starts at $0.572$ fm.
\end{itemize}
On the other hand, the origin of the delay could be the temperature/energy density
dependence of $\hat q$. Thus, we have also studied the case where $\hat q=0$ for $T>T_{\rm cut}=350$ or 380 MeV (see Fig.~\ref{fig:Tcut} in~\ref{appendix-sec1}), which  suppresses  quenching at early times when the energy density is large. 

\paragraph{High-$p_T$ harmonics:}
Up to this point, we have calculated the medium-modified particle spectra, Eq.~\eqref{eq:crossec}, using the method described in Ref.~\cite{Andres:2016iys} but for a hydrodynamic profile produced for a single event. Then we average these single event spectra over all events in a given centrality class to produce the corresponding spectrum for that centrality class.
At this stage, the $K$-factor in Eq.~(\ref{eq:qhat}) can be fitted to the experimental $R_{AA}$ data for a given centrality class. Once the $K$-factor is fixed, the harmonic coefficients associated to the $R_{AA}(p_T,\phi)$ Fourier series $v_n^{hard}$ are calculated in the corresponding  centrality class, event by event. Then, each $v_n^{hard}$ is \textit{correlated} with the soft flow harmonic in the event and, finally, an average over all the events in the centrality class is performed:
\begin{equation}
\resizebox{\linewidth}{!}
{$v_n^{\rm SP}\left(p_T\right) =\frac{\left\langle v_n^{soft} v_n^{hard}\left(p_T\right)\cos\left[n\left(\psi_n^{soft}-\psi_n^{hard}(p_T)\right)\right]\right\rangle}{\sqrt{\left\langle {\left(v_n^{soft}\right)}^2\right\rangle}},$}
\label{eq:vnsp}
\end{equation}
where $\psi_n^{soft}$ is the event plane angle and $\langle ... \rangle$ denotes the average over the events. This is the
scalar product definition of the high-$p_T$ azimuthal harmonics \cite{Luzum:2012da,Noronha-Hostler:2016eow}.

\begin{figure*}[th]
\includegraphics[width=\textwidth]{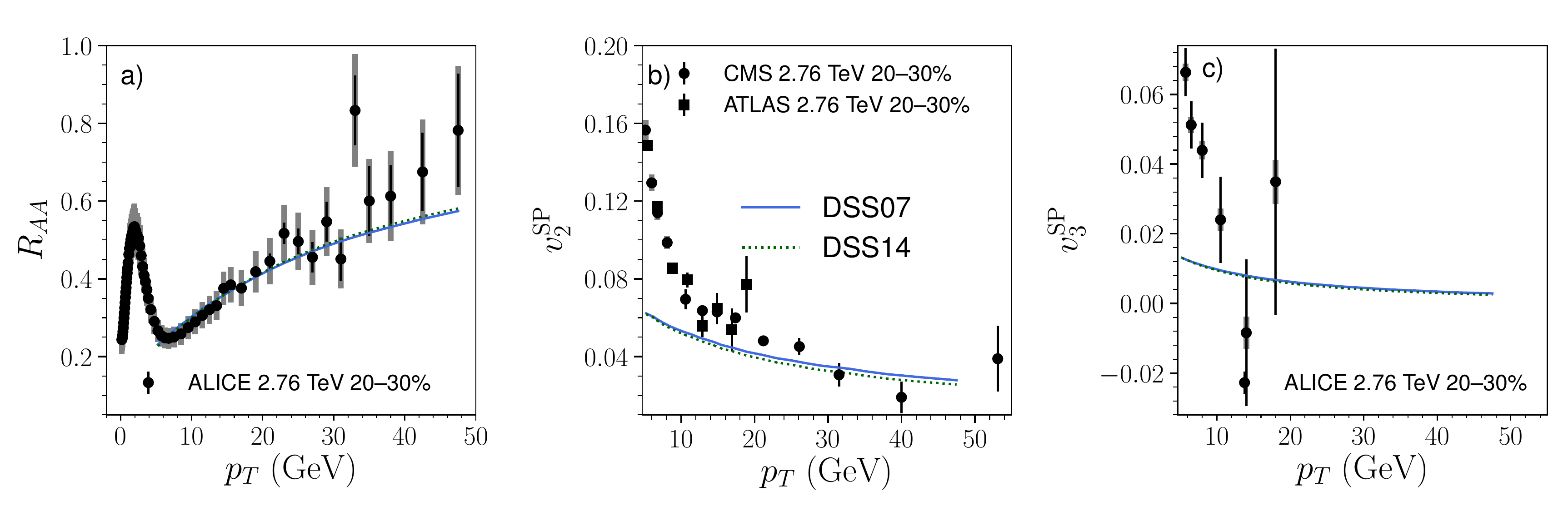}
\vskip -0.4cm
\caption{(Color online) (a) Suppression of inclusive charged particles, (b) high-$p_T$ elliptic flow, (c) high-$p_T$ triangular flow for the 20--30\% centrality class of $\sqrt{s_{\mathrm{NN}}}$ = 2.76 TeV Pb-Pb collisions at the LHC, computed as a function of $p_T$. Experimental data  are from \cite{Abelev:2012hxa,Chatrchyan:2012xq,ATLAS:2011ah,Abelev:2012di}. The blue solid and green dotted lines correspond, respectively, to the use of DSS07  \cite{deFlorian:2007aj} and DSS14 \cite{deFlorian:2014xna} FFs. For the initial and final times of the energy loss,  Case ii)  $\tau_{\rm q}=0.197$ and  $T_{\mathrm{q}} = T_{\mathrm{chem}}$ = 175 MeV are taken.}
\label{fig:ffs}
\end{figure*}

\section{Results} 

\begin{figure*}[th]
\includegraphics[width=\textwidth]{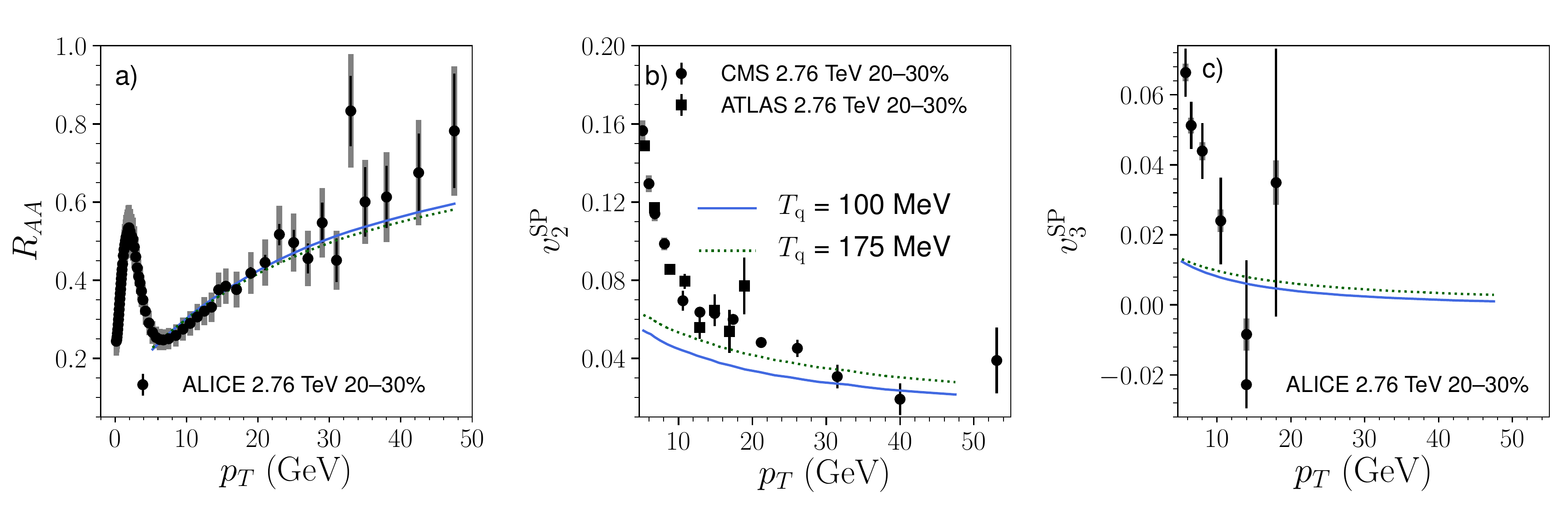}
\vskip -0.4cm
\caption{(Color online) (a) $R_{AA}(p_T)$, (b) $v_2^{\rm SP}(p_T)$, (c) $v_3^{\rm SP}(p_T)$  for the 20--30\% centrality class of $\sqrt{s_{\mathrm{NN}}}$ = 2.76 TeV Pb-Pb collisions at the LHC compared to their respective experimental data \cite{Abelev:2012hxa,Chatrchyan:2012xq,ATLAS:2011ah,Abelev:2012di}. The blue solid line corresponds to stopping the energy loss at the kinetic freeze-out, $T_\mathrm{q} = T_{\mathrm{dec}}$ = 100 MeV. For the green dotted line the quenching finishes at $T_\mathrm{q} = T_{\mathrm{chem}}$ = 175 MeV. DSS07 \cite{deFlorian:2007aj} FFs and Case ii) $\tau_{\rm q}=0.197$ fm are employed.}
\label{fig:Tq}
\end{figure*}

We restrict our study of the nuclear modification factor and the high-$p_T$ harmonics to one center of mass energy and one centrality class:  LHC Pb-Pb 20--30\% semi-central collisions at $\sqrt{s_{\mathrm{NN}}}$ = 2.76 TeV. We have already analyzed the energy and centrality dependence of the nuclear modification factor for several smooth-averaged hydrodynamics in Ref.~\cite{Andres:2016iys}, showing that, surprisingly, the $K$-factor for a given center of mass energy seems to be almost independent of the centrality of the collision. More recently, similar results have been found by all the phenomenological works that set the dependence of the medium parameter on the medium properties to be local and monotonous \cite{Bianchi:2017wpt, Casalderrey-Solana:2018wrw}. Finally, in Ref.~\cite{Andres:2017awo}, we have also checked that using an EbyE formalism, the EKRT hydrodynamic simulation employed also here, the conclusions obtained in Ref.~\cite{Andres:2016iys} remain. 

We compute the nuclear modification factor for a set of values of our free parameter, the $K$-factor, as explained in the previous sections. Next, we perform a $\chi^2$-fit to determine the $K$-value that better describes ALICE $R_{AA}$ data \cite{Abelev:2012hxa} for $p_T > 5$ GeV -- to stay in the pQCD region\footnote{Considering only data with $p_T > 10$ GeV does not modify our main results and conclusions.}. Then, the fitted $K$ is used to obtain the high-$p_T$ asymmetries by means of the scalar product given by Eq.~(\ref{eq:vnsp}). In Fig.~\ref{fig:ffs} we show the dependence of these observables on the FFs employed, i.e., DSS07  or DSS14. In this figure, there is neither energy loss before the initial proper time of the hydrodynamic profile, $\tau_{\rm f} = 0.197$ fm, nor after the chemical freeze-out, $T_{\mathrm{chem}}$ = 175 MeV. It can be seen that, independently of the FFs used, our model fairly-well describes the $R_{AA}$ but underestimates the azimuthal asymmetries in the hard sector. Moreover, our calculations of both the nuclear modification factor and the high-$p_T$ harmonics are hardly sensitive to the FFs. Consequently, any of them can be implemented in our computations, without altering our conclusions. Hereafter, results were obtained using DSS07 FFs.

In Fig.~\ref{fig:Tq} we analyze how the $R_{AA}$ and the jet harmonic coefficients vary with the end-point of the energy loss. As in the previous figure, we assume here no energy loss before the starting time of EKRT hydrodynamic profile, that is, Case ii) $\tau_{\rm q}=0.197$, according to the notation in the preceding section. While the nuclear modification factor can be well described both with and without energy loss in the hadronic phase, the high-$p_T$ asymmetries are sensitive, especially the $v_2^{\rm SP}(p_T)$, to the end-point of the quenching, pointing out to a better description of the data when there is only energy loss in the partonic phase. Nevertheless, no matter when we stop our simulation, yet the jet harmonic coefficients remain underestimated.
\begin{figure*}[th]
\includegraphics[width=\textwidth]{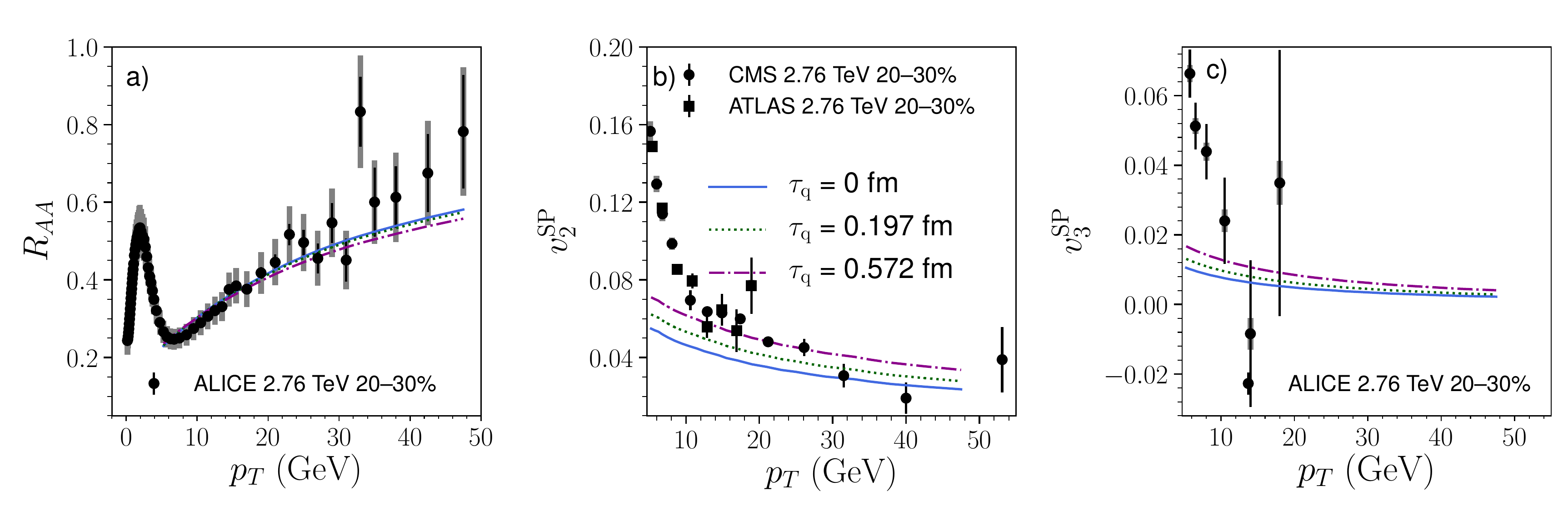}
\vskip -0.4cm
\caption{(Color online) (a) $R_{AA}(p_T)$, (b) $v_2^{\rm SP}(p_T)$, (c) $v_3^{\rm SP}(p_T)$ for the 20--30\% centrality class of  $\sqrt{s_{\mathrm{NN}}}$ = 2.76 TeV Pb-Pb collisions at the LHC compared to their respective experimental data \cite{Abelev:2012hxa,Chatrchyan:2012xq,ATLAS:2011ah,Abelev:2012di}. The blue solid, $\tau_{\rm q}=0$ fm, dotted green, $\tau_{\rm q}=0.197$ fm, and dashed-dotted purple, $\tau_{\rm q}=0.572$ fm, lines correspond, respectively, to Cases i), ii) and iii) of the early times treatment. DSS07 \cite{deFlorian:2007aj} FFs and $T_{\mathrm{q}} = T_{\mathrm{chem}}$ = 175 MeV are used.}
\label{fig:tauq}
\end{figure*}

The dependence of the $R_{AA}(p_T)$,  $v_2^{\rm SP}(p_T)$, and $v_3^{\rm SP}(p_T)$ on the starting time on the energy loss is presented in Fig.~\ref{fig:tauq}. This is done for the case where there is no quenching in the hadronic phase, $T_{\mathrm{q}} = T_{\mathrm{chem}}$. As it can be seen on the left panel of this figure, the dependence of the nuclear modification factor on $\tau_{\rm q}$ is mild, however, the corresponding $K$-fitted values for the three curves of this panel, shown in Table~\ref{table:kfactor}, are quite different. Regarding the asymmetries in the hard sector, Fig.~\ref{fig:tauq} shows that they are very sensitive to the starting point of the quenching. Actually, the high-$p_T$ $v_2$ experimental data can be described substantially better within our formalism if and only if the starting point of the energy loss is delayed up to $\sim 0.6$ fm {-- the corresponding $\chi^2/{\rm d.o.f.}$ are shown in Table~\ref{table:kfactor}. This corresponds to the set-up employed in any approach that aims to describe the jet harmonics coefficients using a smooth dependence of the medium parameter on the medium properties \cite{Noronha-Hostler:2016eow, Betz:2016ayq,Zigic:2018ovr}.

\begin{table}[h]
\begin{center}
\resizebox{\linewidth}{!}{%
\begin{tabular}{ c c c}
  \hline  
  Early time extrapolation & $K$-factor &  $\chi^2/{\rm d.o.f.}$ for  $v_2$\\
  \hline  
    Case i) $\tau_{\rm q}=0$ fm & $2.120^{+0.091}_{-0.074}$& 26.2\\
 \vspace{0.7mm}
 Case ii) $\tau_{\rm q}=0.197$ fm & $2.90^{+0.13} _{-0.11} $& 12.9\\   
  \vspace{0.7mm}
 Case iii) $\tau_{\rm q}=0.572$ fm & 4.56 $\pm$ 0.20 & 3.5\\
   \hline
\end{tabular}}
\vskip -0.1cm
\caption{$K$-factor obtained from fits to the ALICE $R_{AA}$ data \cite{Abelev:2012hxa} for the three different early time extrapolations and the corresponding $\chi^2/{\rm d.o.f.}$ for the $v_2$ CMS data with $p_T>10$ GeV.
DSS07 FFs and $T_{\mathrm{q}} = T_{\mathrm{chem}} =$ 175 MeV are employed.}
\label{table:kfactor}
\end{center}
\end{table}

\section{Conclusions} 
In this Letter we have computed the nuclear modification factor and the high-$p_T$ harmonics $v_2$, $v_3$ for charged particle production in  20--30\% centrality class $\sqrt{s_{\mathrm{NN}}}$ = 2.76 TeV Pb-Pb collisions at the LHC. The calculations
are done by using the formalism of QWs embedded in the state-of-the art EbyE EKRT hydrodynamic model of the medium. We have analyzed the dependence of these observables on the FFs, on the lack - or not - of energy loss in the hadronic phase of the evolution, and on the starting time of the quenching. Any work that correctly determines the inclusive particle suppression and harmonic coefficients in the hard sector starts the energy loss at the initial time of the hydrodynamic simulation employed, which usually is $\tau_{\rm f}$ = 0.6 fm (or later). Therefore, they implicitly assume no quenching during the first 0.6 fm after the collision. Since the starting time of the EKRT hydrodynamic evolution is $\tau_{\rm f}$ = 0.197 fm, it provides the first framework that enables the variation of the quenching in the early stages of the evolution, and thus the determination of its beginning in a controlled way. We find that the simultaneous and proper description of these three observables requires no energy loss for the first $\sim 0.6$ fm after the collision (or at large $T>350$ MeV), in accord with the implicit set-up in other studies.

Clearly, our result comes from a smaller $\hat q$ at early times, but
we lack a conclusive physical explanation for this finding. It would be tempting to link $\hat q$ with the Knudsen number which is large at these early times. For instance, in weakly coupled theories $\hat q /T^3 \propto (\eta/s)^{-1}$ \cite{Majumder:2007zh}. Therefore, a large Knudsen number due to a large $\eta/s$ (and not due to large gradients) would imply a small $\hat q$ and the suppression of jet quenching. We also note that, although the EoS affects the temperature dependence of $\hat q$ through Eq.~(\ref{eq:qhat}) to some extent, the high temperature part of the EoS is very well established from lattice QCD calculations~\cite{Ratti:2018ksb}. On the other hand, the low-temperature part of the EoS~\cite{Hirano:2002ds} can be strongly affected by the chemical freeze-out. However, we have tested, by changing the quenching endpoint, that the hadronic evolution does not alter our conclusions.

We conclude that this is not a particular feature of our approach but a general outcome. Hence, high-$p_T$ asymmetries are introduced here, for the first time, as a direct signature of the less known initial stages of the collision, showing the impossibility of the simultaneous description of the experimental measurements on the charged hadron suppression and the azimuthal asymmetries without strongly suppressing the energy loss for the first $\sim 0.6$ fm after the collision. This work clearly shows that exploiting the versatility of jet quenching to access different time-scales offers unique possibilities to improve our understanding of the initial stages in heavy-ion collisions,
and is extendable from large to small systems.

\section*{Acknowledgements}
We acknowledge helpful discussions with J. Noronha-Hostler, computational resources from the CSC-IT Center for Science in Espoo, Finland, and financial support  by the US DOE (CA, contract DEAC05-06OR23177 under which Jefferson Science Associates, LLC operates Jefferson Lab), the Academy of Finland (HN, project 297058), the ERC (RP, grant no. 725369), MICINN of Spain (NA,CAS, project FPA2017-83814-P and Unidad de Excelencia Mar\'{\i}a de Maetzu  MDM-2016-069), Xunta de Galicia (NA,CAS, Conseller\'{\i}a de Educaci\'on) and FEDER (NA,CAS). This work has been performed within COST Action CA15213 THOR.

\appendix
\section{Additional checks}
\label{appendix-sec1}
\renewcommand\thefigure{A.\arabic{figure}}    
\setcounter{figure}{0}    

\paragraph{Different centralities:} We have investigated the effect of the cut in time for different centrality classes.  The results for $R_{AA}(p_T)$ and $v_2^{\rm SP}(p_T)$ for the 0--10\% and 40--50\% centrality classes of $\sqrt{s_{\mathrm{NN}}}$ = 2.76 TeV Pb-Pb collisions at the LHC are shown, respectively, in Fig.~\ref{fig:T0-10} and Fig.~\ref{fig:T40-50}.  For both centrality classes, we consider again the three early times extrapolations: $\tau_{\rm q}=0$ fm, $\tau_{\rm q}=0.197$ fm and $\tau_{\rm q}=0.572$ fm, taking DSS07 \cite{deFlorian:2007aj} FFs and $T_{\mathrm{q}} = T_{\mathrm{chem}}$ = 175 MeV. The corresponding central values of the $K$-factor are, respectively, 2.12, 2.79 and 4.12 for the 0--10\% centrality class and 2.14, 3.10 and 5.27 for the 40--50\% centrality class, in line with the findings in \cite{Andres:2016iys}. The improvement in the description of $v_2$ with increasing $\tau_q$ is manifest.

\paragraph{Energy loss modeling:} We have examined the effect of using a different energy loss model. Within the same formalism of the QWs, we have changed the approximation used to compute the radiation spectrum from multiple soft scatterings to a single hard scattering, that is, the $N=1$ opacity limit (taking $\bar R=R/3$ and $\bar \omega_c=\omega_c/3$, see \cite{Salgado:2003gb} and also \cite{Armesto:2011ht}). Note that the perturbative tails largely differ between these two approximations. We show in Fig.~\ref{fig:GLV} the results for $R_{AA}(p_T)$ and $v_2^{\rm SP}(p_T)$ for the 20--30\% centrality class of $\sqrt{s_{\mathrm{NN}}}$ = 2.76 TeV Pb-Pb collisions at the LHC in the single opacity approximation, together with the ones in the multiple soft scattering approximation  for $\tau_{\rm q}=0$ fm, $\tau_{\rm q}=0.197$ fm and $\tau_{\rm q}=0.572$ fm  (using DSS07 \cite{deFlorian:2007aj} FFs and $T_{\mathrm{q}} = T_{\mathrm{chem}}$ = 175 MeV). The corresponding central values of the $K$-factor for the the $N=1$ opacity curves are 2.80, 3.80 and 6.03, respectively. While the transverse momentum dependence of the results is somewhat different, the improvement in the description of $v_2$ with increasing $\tau_q$ is evident.

\begin{figure*}[th]
\includegraphics[width=\textwidth]{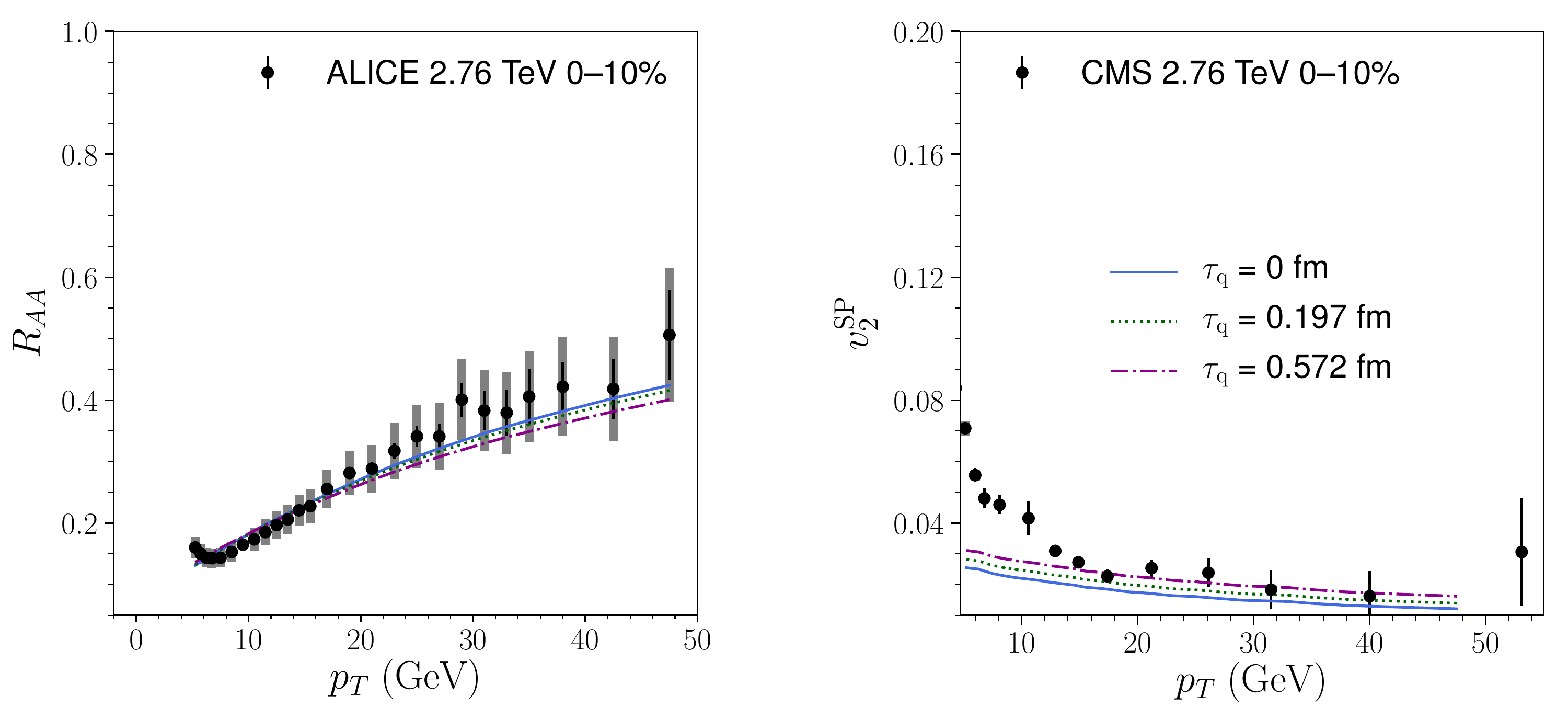}
\vskip -0.4cm
\caption{(Color online) (Left) $R_{AA}(p_T)$, (right) $v_2^{\rm SP}(p_T)$ for the 0--10\% centrality class of  $\sqrt{s_{\mathrm{NN}}}$ = 2.76 TeV Pb-Pb collisions at the LHC compared to their respective experimental data \cite{Abelev:2012hxa,Chatrchyan:2012xq}. The blue solid, $\tau_{\rm q}=0$ fm, dotted green, $\tau_{\rm q}=0.197$ fm, and dashed-dotted purple, $\tau_{\rm q}=0.572$ fm, lines correspond, respectively, to Cases i), ii) and iii) of the early times treatment. DSS07 \cite{deFlorian:2007aj} FFs and $T_{\mathrm{q}} = T_{\mathrm{chem}}$ = 175 MeV are used.}
\label{fig:T0-10}
\end{figure*}

\begin{figure*}[h!]
\includegraphics[width=\textwidth]{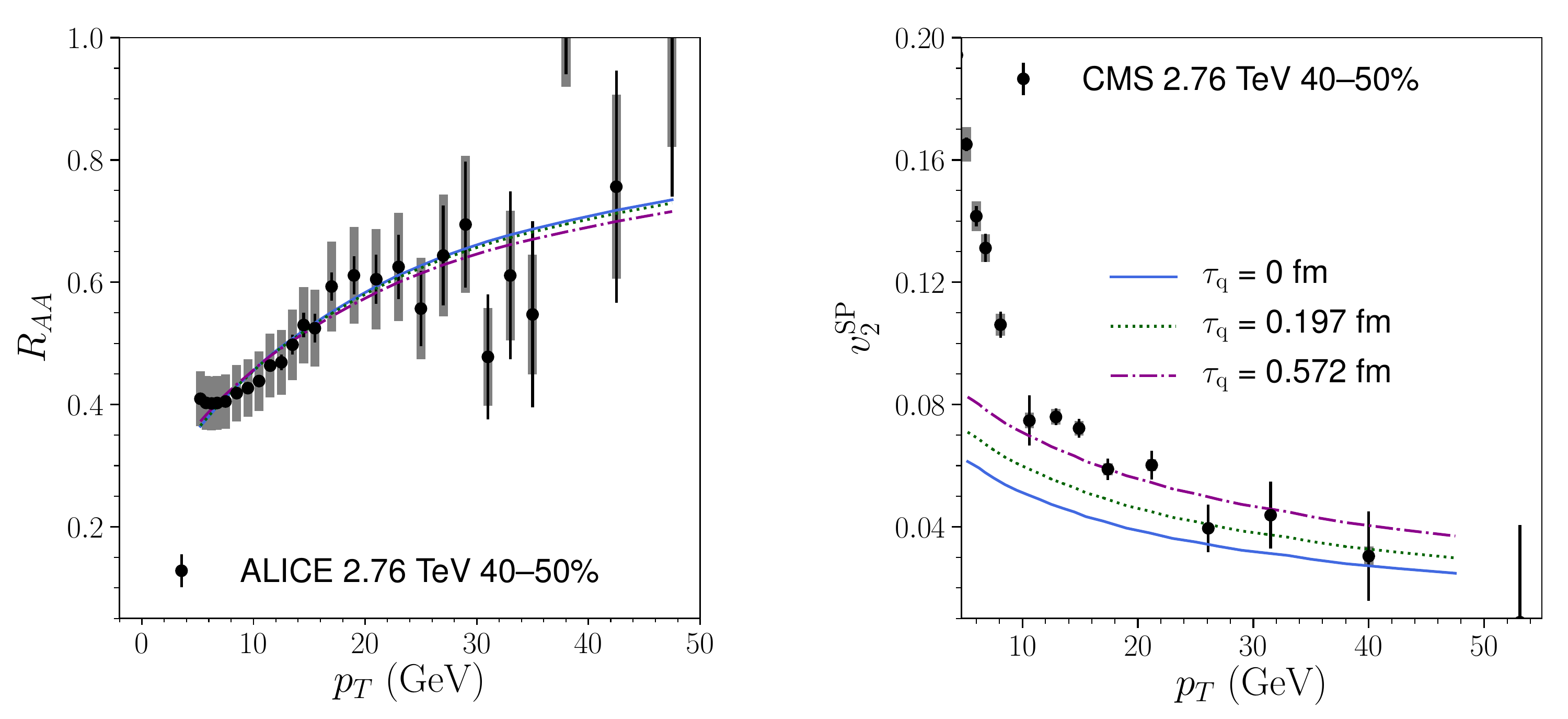}
\vskip -0.4cm
\caption{(Color online) (Left) $R_{AA}(p_T)$, (right) $v_2^{\rm SP}(p_T)$ for the 40--50\% centrality class of  $\sqrt{s_{\mathrm{NN}}}$ = 2.76 TeV Pb-Pb collisions at the LHC compared to their respective experimental data \cite{Abelev:2012hxa,Chatrchyan:2012xq}. The blue solid, $\tau_{\rm q}=0$ fm, dotted green, $\tau_{\rm q}=0.197$ fm, and dashed-dotted purple, $\tau_{\rm q}=0.572$ fm, lines correspond, respectively, to Cases i), ii) and iii) of the early times treatment. DSS07 \cite{deFlorian:2007aj} FFs and $T_{\mathrm{q}} = T_{\mathrm{chem}}$ = 175 MeV are used.}
\label{fig:T40-50}
\end{figure*}

\paragraph{Cuts in temperature:} We have evaluated the possibility of a different way of cutting the quenching at the initial stages of the collision. Specifically, we have taken $\hat q=0$ for $T>T_{\rm cut}=350$ or 380 MeV\footnote{Using DSS07 \cite{deFlorian:2007aj} FFs and ending the energy loss at $T_{\mathrm{q}} = T_{\mathrm{chem}}$ = 175 MeV.}. The corresponding central values for the $K$-factor are 7.00 and 5.77 for $T_{\rm cut}=350$ and 380 MeV respectively. The results for  $R_{AA}(p_T)$ and $v_2^{\rm SP}(p_T)$ for the 20--30\% centrality class of $\sqrt{s_{\mathrm{NN}}}$ = 2.76 TeV Pb-Pb collisions at the LHC are shown in Fig.~\ref{fig:Tcut}, together with the results with $\tau_q=0.572$ fm and no cut in temperature. It turns out that the effect of decreasing $T_{\rm cut}$ is similar to that of increasing $\tau_q$, as expected.

\begin{figure*}[h!]
\includegraphics[width=\textwidth]{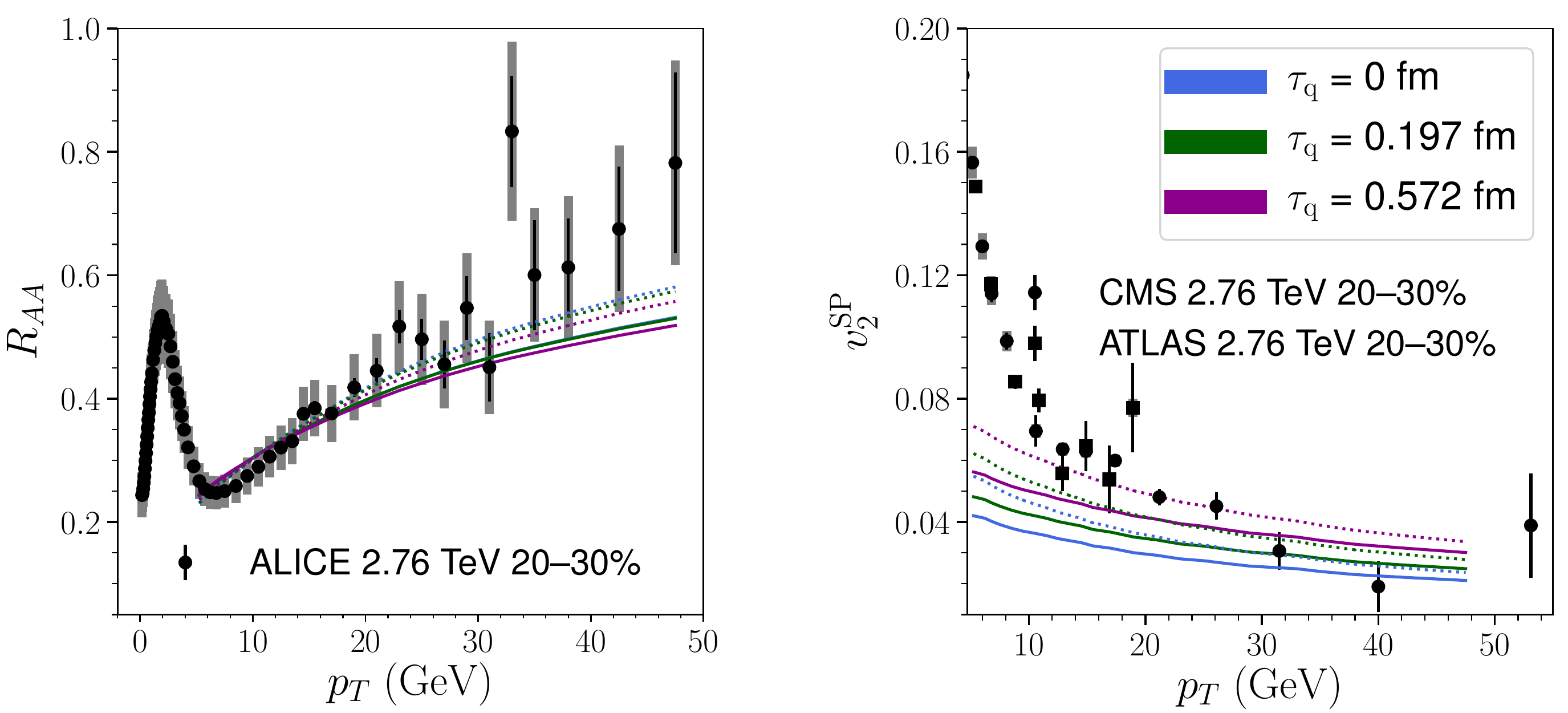}
\vskip -0.4cm
\caption{(Color online) (Left) $R_{AA}(p_T)$, (right) $v_2^{\rm SP}(p_T)$  for the 20--30\% centrality class of $\sqrt{s_{\mathrm{NN}}}$ = 2.76 TeV Pb-Pb collisions at the LHC compared to their respective experimental data \cite{Abelev:2012hxa,Chatrchyan:2012xq,ATLAS:2011ah}. The color of the lines correspond to the early times treatment employed, that is, blue for Case i) $\tau_{\rm q}=0$ fm, green for Case ii) $\tau_{\rm q}=0.197$ fm and purple for Case iii) $\tau_{\rm q}=0.572$ fm. Solid lines refer to the results in the single opacity approximation, while dotted lines correspond to the multiple soft scattering approximation used in the main part of the work, that is, Figs.~\ref{fig:ffs}, \ref{fig:Tq} and \ref{fig:tauq}, and in all other the Figs.~in this Appendix. DSS07 \cite{deFlorian:2007aj} FFs and $T_{\mathrm{q}} = T_{\mathrm{chem}}$ = 175 MeV are used.}
\label{fig:GLV}
\end{figure*}

\begin{figure*}[h!]
\includegraphics[width=\textwidth]{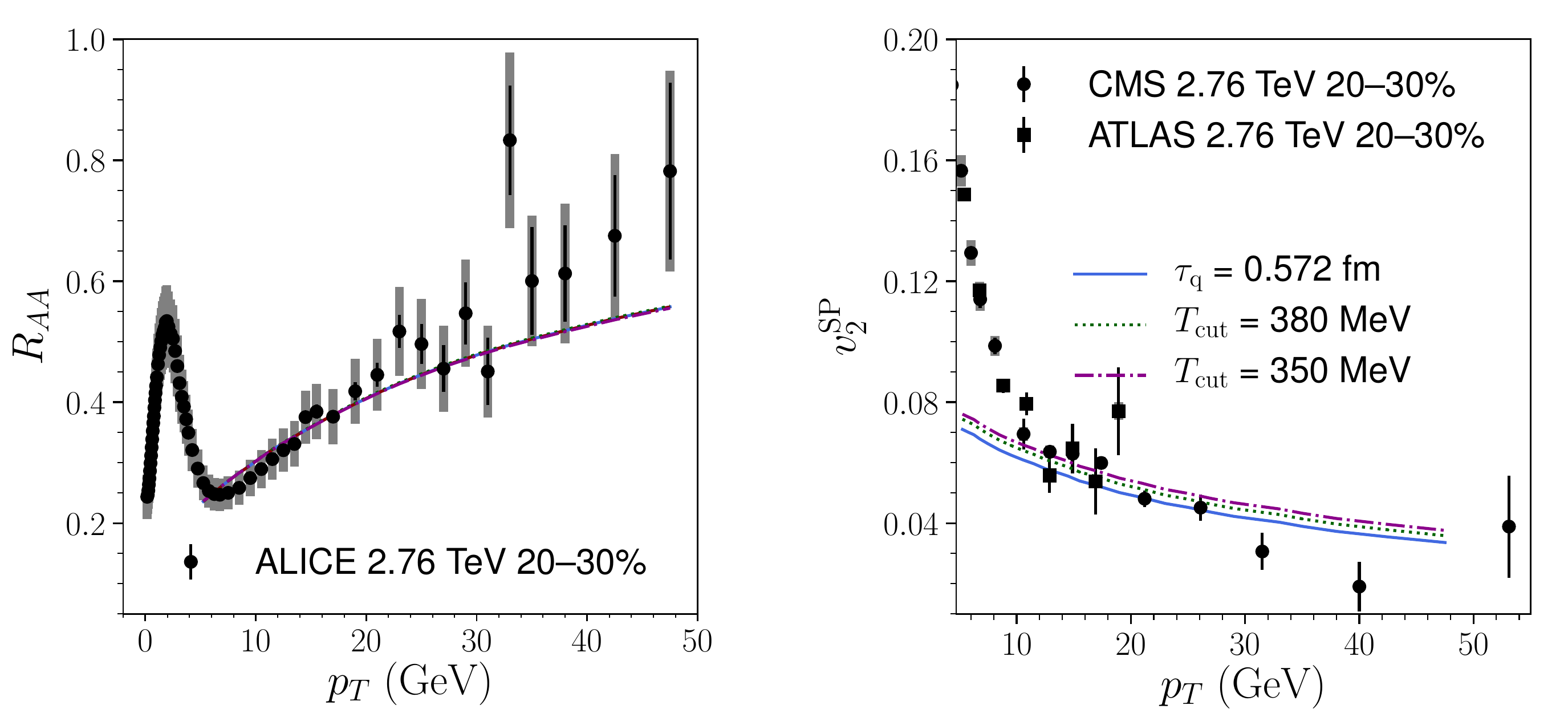}
\vskip -0.4cm
\caption{(Color online) (Left) $R_{AA}(p_T)$, (right) $v_2^{\rm SP}(p_T)$  for the 20--30\% centrality class of $\sqrt{s_{\mathrm{NN}}}$ = 2.76 TeV Pb-Pb collisions at the LHC compared to their respective experimental data \cite{Abelev:2012hxa,Chatrchyan:2012xq,ATLAS:2011ah}. The solid blue line corresponds to a cut in time with $\tau_{\rm q}=0.572$ fm. The dotted green and dashed-dotted purple lines correspond to cuts in temperature $T_{\rm cut}=380$ and 350 MeV, respectively. DSS07 \cite{deFlorian:2007aj} FFs and $T_{\mathrm{q}} = T_{\mathrm{chem}}$ = 175 MeV are used.}
\label{fig:Tcut}
\end{figure*}

\bibliography{mybibfile.bib}

\end{document}